\documentclass[12pt]{article}

\RequirePackage{cite}
\RequirePackage{times}
\RequirePackage{fullpage}
\RequirePackage{ifthen}

\makeatletter 
\def\@cite#1#2{$^{\mbox{\scriptsize #1\if@tempswa , #2\fi}}$}
\makeatother

\newcommand{\spacing}[1]{\renewcommand{\baselinestretch}{#1}\large\normalsize}
\spacing{2}

\def\@maketitle{%
  \newpage\spacing{1}\setlength{\parskip}{12pt}%
    {\Large\bfseries\noindent\sloppy \textsf{\@title} \par}%
    {\noindent\sloppy \@author}%
}

\newenvironment{affiliations}{%
    \setcounter{enumi}{1}%
    \setlength{\parindent}{0in}%
    \slshape\sloppy%
    \begin{list}{\upshape$^{\arabic{enumi}}$}{%
        \usecounter{enumi}%
        \setlength{\leftmargin}{0in}%
        \setlength{\topsep}{0in}%
        \setlength{\labelsep}{0in}%
        \setlength{\labelwidth}{0in}%
        \setlength{\listparindent}{0in}%
        \setlength{\itemsep}{0ex}%
        \setlength{\parsep}{0in}%
        }
    }{\end{list}\par\vspace{12pt}}

\renewenvironment{abstract}{%
    \setlength{\parindent}{0in}%
    \setlength{\parskip}{0in}%
    \bfseries%
    }{\par\vspace{-4pt}}

\usepackage{times}

\usepackage{graphicx}

\def \aap{Astron.~Astrophys.}
\def \apjl{Astrophys.~J.}
\def \apjs{Astrophys.~J.~Supp.}
\def \apj{Astrophys.~J.}

\def \mnras{Mon.~Not.~R.~Astron.~Soc.}
\def \pasp{PASP}

\def\dg{^{\circ}}

\newcommand{\lsim}{\,\rlap{\raise 0.35ex\hbox{$<$}}{\lower 0.7ex\hbox{$\sim$}}\,}
\newcommand{\gsim}{\,\rlap{\raise 0.35ex\hbox{$>$}}{\lower 0.7ex\hbox{$\sim$}}\,}

\newcommand{\la}{skywalker@galaxy.far.far.away}

\def\la{\mathrel{\hbox{\rlap{\hbox{\lower4pt\hbox{$\sim$}}}\hbox{$<$}}}}
\def\ga{\mathrel{\hbox{\rlap{\hbox{\lower4pt\hbox{$\sim$}}}\hbox{$>$}}}}

\def\kms {\hbox{km\,s$^{-1}$}}
\def\ergs {\hbox{erg\,s$^{-1}$}}
\newcommand{\e}[1]{\times 10^{#1}}

\newcommand{\arcsec}{${}^{\prime \prime}$}
\newcommand{\msun}{M$_\odot$}
\newcommand{\wll}{$\lambda \lambda$}

\begin{document}

\noindent
 {\Large {\bf {\fontfamily{phv}\selectfont X-ray illumination of the ejecta of Supernova 1987A }}}

\noindent
J.~Larsson$^1$, 
C.~Fransson$^1$, 
G.~\"Ostlin$^1$, 
P.~Gr\"oningsson$^1$, 
A.~Jerkstrand$^1$,
C.~Kozma$^1$, 
J.~Sollerman$^1$, 
P.~Challis$^2$, 
R.~P.~Kirshner$^2$, 
R.~A.~Chevalier$^3$, 
K.~Heng$^{4}$, 
R.~McCray$^5$, 
N.~B.~Suntzeff$^6$, 
P.~Bouchet$^7$,
A.~Crotts$^8$,
J.~Danziger$^{9}$,
E.~Dwek$^{10}$,
K.~France$^{11}$,
P.~M.~Garnavich$^{12}$,
S.~S.~Lawrence$^{13}$,
B.~Leibundgut$^{14}$,
P.~Lundqvist,$^1$
N.~Panagia$^{15,16,17}$,
C.~S.~J.~Pun$^{18}$,
N.~Smith$^{19}$,
G.~Sonneborn$^{10}$,
L.~Wang$^{20}$,
J.~C.~Wheeler$^{21}$
\begin{affiliations}

\item Department of Astronomy, The Oskar Klein Centre, Stockholm University, 106 91 Stockholm, Sweden.
\item Harvard-Smithsonian Center for Astrophysics, 60 Garden Street, MS-19, Cambridge, MA 02138, USA.
\item Department of Astronomy, University of Virginia, Charlottesville, VA 22903, USA 
\item Eidgen\"ossische Technische Hochschule Z\"urich, Institute for Astronomy, Wolfgang-Pauli-Strasse 27, CH-8093, Z\"urich, Switzerland.
\item JILA, University of Colorado, Boulder, CO 803090-0440, USA.
\item George P. and Cynthia Woods Mitchell Institute for Fundamental Physics and Astronomy, Texas A\&M University, Department of Physics and Astronomy, 
College Station, TX 77843, USA. 
\item DSM/IRFU/Service d\'Astrophysique Commissariat \'a l\'Energie Atomique et aux \'Energies Alternatives, Saclay, Orme des Merisiers, FR 91191 Gif-sur-Yvette, France.
\item Department of Astronomy, Mail Code 5240, Columbia University, 550 West 120th Street, New 
York, NY 10027, USA.
\item Osservatorio Astronomico di Trieste, Via Tiepolo 11, Trieste, 34131, Italy. 
\item NASA Goddard Space Flight Center, Code 665, Greenbelt, MD 20771, USA. 
\item Center for Astrophysics and Space Astronomy, University of Colorado, Boulder, CO 80309, USA.
\item 225 Nieuwland Science, University of Notre Dame, Notre Dame, IN 46556-5670, USA.
\item Department of Physics and Astronomy, Hofstra University, Hempstead, NY 11549, USA. 
\item ESO, Karl-Schwarzschild-Strasse 2, 85748 Garching, Germany. 
\item Space Telescope Science Institute, 3700 San Martin Drive, Baltimore, MD 21218, USA.
\item INAF-CT, Osservatorio Astroﬁsico di Catania, Via S. Soﬁa 78, I-95123 Catania, Italy. 
\item Supernova Limited, OYV \#131, Northsound Road, Virgin Gorda, British Virgin Islands. 
\item Department of Physics, University of Hong Kong, Pok Fu Lam Road, Hong Kong, China. 
\item Steward Observatory, University of Arizona, 933 North Cherry Avenue, Tucson, AZ 85721, USA.
\item Department of Physics and Astronomy, Texas A\&M University, College Station, TX 77843-4242, USA. 
\item Department of Astronomy, University of Texas, Austin, TX 78712-0259, USA.

\end{affiliations}

\begin{abstract}

When a massive star explodes as a supernova, substantial amounts of
radioactive elements---primarily ${}^{56}$Ni, ${}^{57}$Ni and
${}^{44}$Ti---are produced\cite{Woosley2002}. After the initial flash
of light from shock heating, the fading light emitted by the supernova
is due to the decay of these elements\cite{Fransson2002}. However,
after decades, the energy for a supernova remnant comes from the shock
interaction between the ejecta and the surrounding
medium\cite{McKee2001}. The transition to this phase has hitherto not
been observed: supernovae occur too infrequently in the Milky Way to
provide a young example, and extragalactic supernovae are generally
too faint and too small. Here we report observations that show this
transition in the supernova SN 1987A in the Large Magellanic
Cloud. From 1994 to 2001 the ejecta faded owing to radioactive decay
of ${}^{44}$Ti as predicted. Then the flux started to increase, more
than doubling by the end of 2009.  We show that this increase is the
result of heat deposited by X-rays produced as the ejecta interacts
with the surrounding material. In time, the X-rays will penetrate
farther into the ejecta, enabling us to analyse the structure and
chemistry of the vanished star.

\end{abstract}

\begin{figure}[h!]
\begin{center}
\resizebox{54mm}{!}{\includegraphics{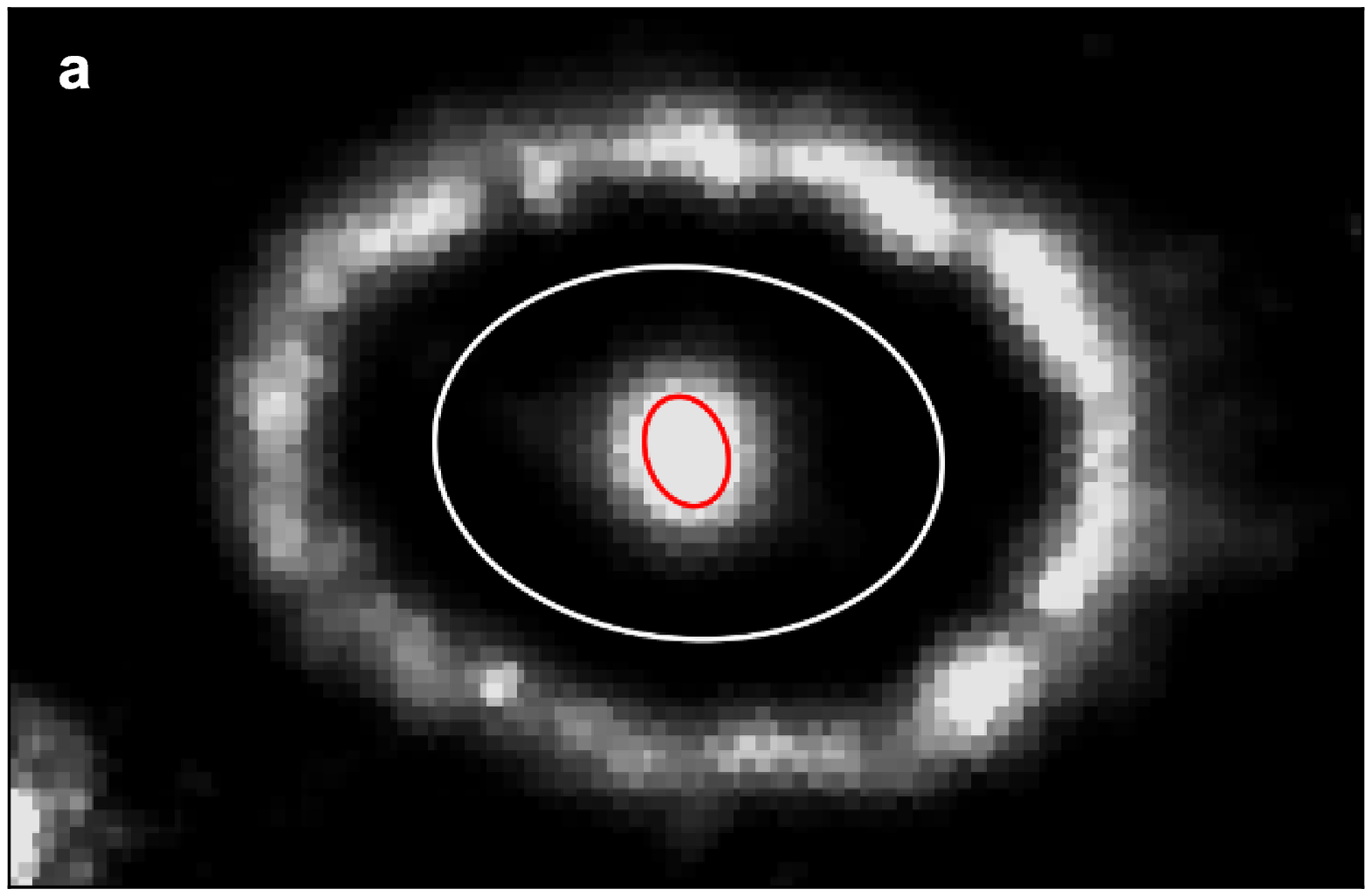}}
\resizebox{54mm}{!}{\includegraphics{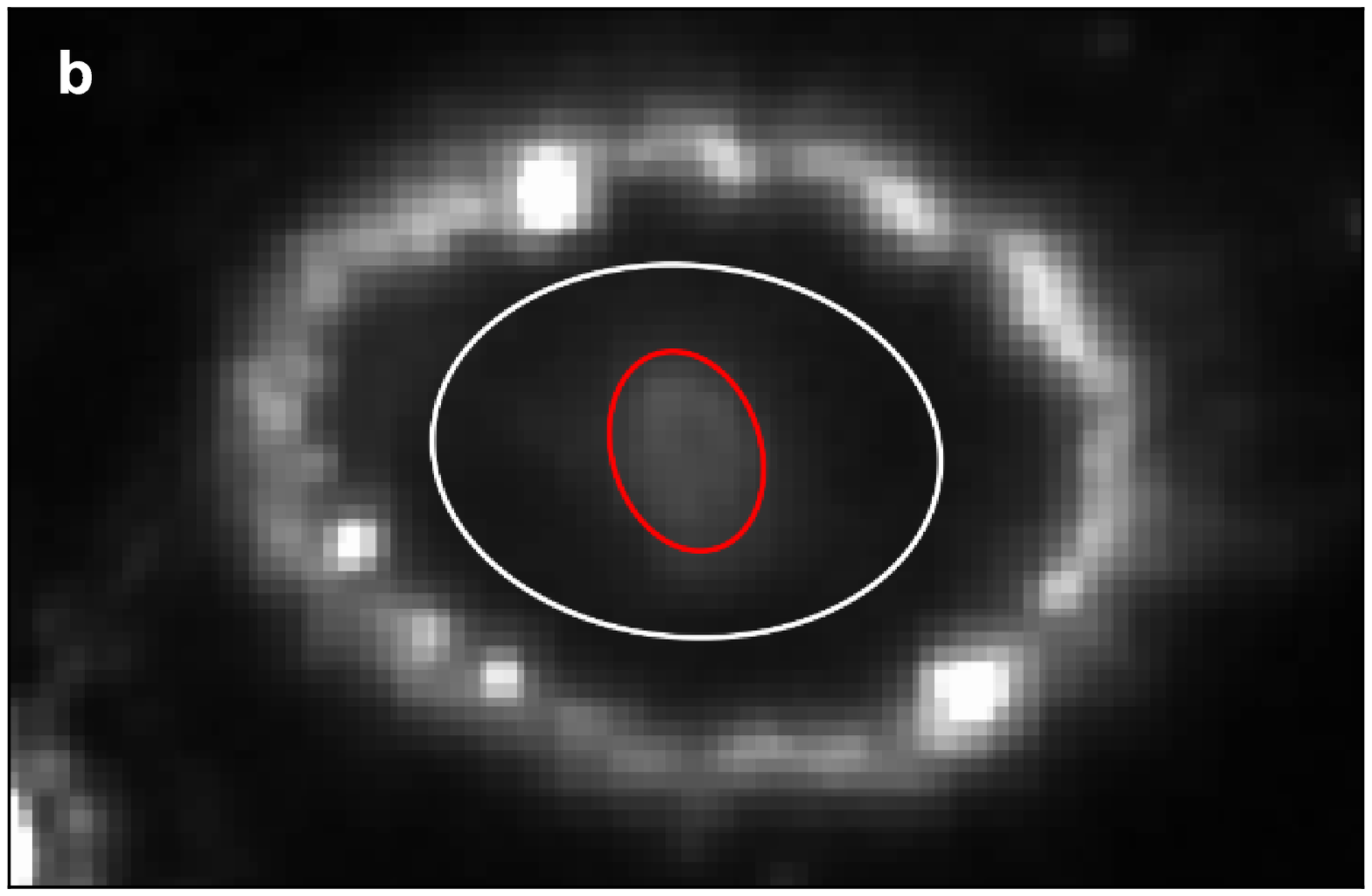}}
\resizebox{54mm}{!}{\includegraphics{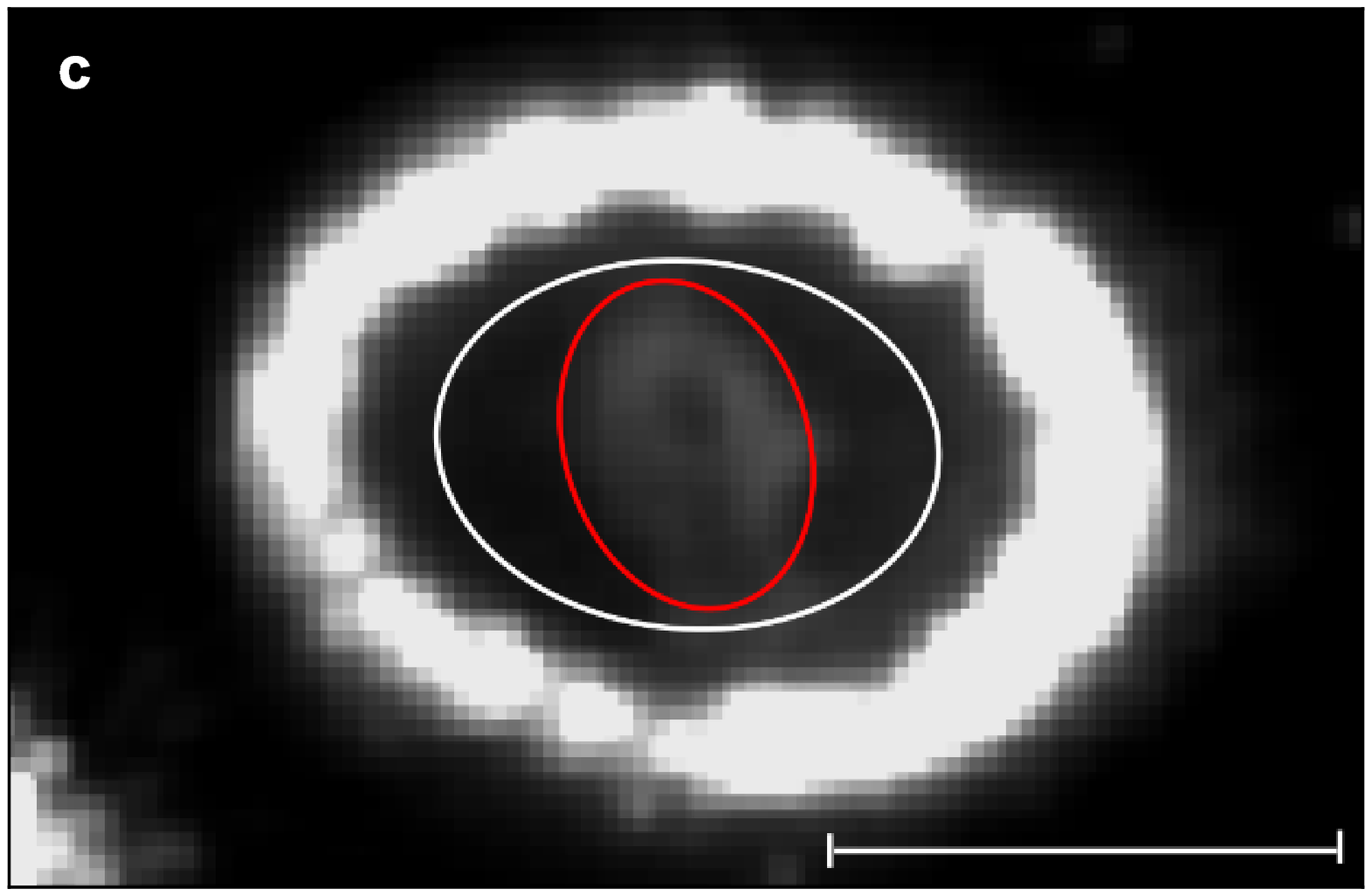}}
\caption{A selection of HST R-band images. The observing dates are
  1994-09-24 ({\bf a}), 2000-11-13 ({\bf b}) and 2009-04-29 ({\bf c}),
  which correspond to 2,770, 5,012 and 8,101 days after the explosion,
  respectively. The scale bar in {\bf c} represents 1$''$. The
  circumstellar ring is inclined at an angle of $45\dg$ with respect
  to the line of sight and is approximately 1.3~light years
  across. The red (inner) ellipse shows the expanding aperture used
  for the light curve in Fig.~\ref{fig2}. By using an initial
  semi-major axis of $0.11''$ for the observation in 1994 we always
  follow the bright, central part of the ejecta without being
  significantly affected by emission from the circumstellar ring. The
  white (outer) ellipse shows the fixed aperture used for one of the
  light curves in Supplementary Fig.~2. The R-band emission from the
  ejecta is dominated by H$\alpha$ emission with a small contribution
  from [Ca I-II] lines, whereas the B-band (Supplementary Fig.~1) is
  dominated by H I and Fe I-II
  lines\cite{Chugai1997,Jerkstrand2011}. Only the densest,
  central parts of the ejecta are visible due to the low surface
  brightness of the outer parts. In reality, the ejecta extend to the
  ring, as is evident from the strong interaction with the ring. }
\label{fig1}
\end{center}
\end{figure}

Due to the proximity of SN~1987A (located only $160,000$ light years
away), we can study the evolution of the supernova (SN) in great
detail. The central ejecta are surrounded by a ring of circumstellar
material (Fig.~\ref{fig1}) that was shed from the star $20,000$ years
before the explosion in 1987\cite{Morris2007}. Since the explosion,
the ejecta have been expanding, and now the outer parts of the ejecta
are colliding with the ring, causing it to brighten at all
wavelengths\cite{2008bGroningsson,Racusin2009,Zanardo2010,Dwek2010}. The
dense, central part of the ejecta contains most of the mass from the
disrupted star and acts as a calorimeter for the energy input to the
SN. We have determined the energy input by tracking the energy output
with the Hubble Space Telescope (HST).

Because the ejecta are roughly elliptical in projection on the sky we
use an elliptical aperture to measure the brightness. To monitor a
constant mass of expanding material, we allowed the measuring aperture
to expand linearly with time. The axes of the aperture were therefore
three times larger in 2009 than in 1994 (Fig.~\ref{fig1}). Using this
aperture, we determined the R-and and B-band light curves of the
ejecta, as shown in Fig.~\ref{fig2} (see Supplementary Table 1 and
Supplementary Information, section 1 for further details of the
observations and light curves). Our measurements show that the flux
from the ejecta decays during the first $\sim5,000$ days after the
explosion, as expected from radioactive input, but then starts to
increase, reaching a level that is 2-3 times higher around day $8,000$
(end of 2009). A new energy source must be present in addition to
radioactive decay.  Below, we consider a model for the declining phase
and then discuss the new energy source that is responsible for the
observed increase in flux.


The energy input to the declining phase of the light curve after $\sim
1,500$ days is expected to come from positrons produced in the decay of
${}^{44}$Ti\cite{Timmes1996,Diehl1998,Fransson2002}. To test this we
use a model\cite{Fransson2002} with abundances taken from the 14E1
explosion model\cite{Shigeyama1990} and a ${}^{44}$Ti mass of $1.4
\times 10^{-4}$ \msun \cite{Jerkstrand2011} (Supplementary
Information, section 3). The model is shown in Fig.~\ref{fig4}
together with the observed broadband luminosities. The good agreement
with the observations up to day $5,000$ confirms that the ${}^{44}$Ti
positrons provide the energy input up to this point. However, after
day $5,000$ the model clearly fails to describe the light curve;
radioactive decay cannot explain the increase in flux that we observe.


One possible origin for the flux increase is the reverse shock that
results from the interaction between the ejecta and the H II region
inside the ring\cite{Michael2003,Smith2005,Heng2006,France2010}. The
reverse shock produces strong Ly$\alpha$ and H$\alpha$ emission, which
increased by a factor of $\sim 1.7$ between 2004 and
2010\cite{France2010}. Although most of this emission originates close
to the ring, there is also a component of projected high-velocity
H$\alpha$-emission that can be traced to the central parts of the
ejecta\cite{France2010} and which would therefore contribute to the
flux we measure. To determine the contribution of the reverse shock to
our light curves we have examined HST STIS spectra from 2004 and 2010
(Supplementary Information, section 2 and Supplementary Fig.~5). The
reverse shock can be isolated in the spectra because of its boxy line
profile, allowing us to place a limit on its contribution at $\la 20
\%$. Furthermore, this changes only marginally between 2004 and 2010,
as the expanding measuring aperture remains well inside the area where
most of the shock emission is seen. Importantly, an increase in flux
is also seen in the [Ca II] doublet lines at rest wavelengths
7,292~\AA\ and 7,324~\AA\ between 2000 -- 2010 (determined from UVES
observations at the ESO/VLT, Fig. \ref{fig2}). These lines have speeds
of $\la 5,000$ \kms, implying that they originate in the inner ejecta
(the projected ejecta speed near the edge of the ring is $\ga 7,000$
\kms \ at the present time). We conclude that the increase in flux
occurs primarily in the inner ejecta and cannot be explained by
emission from the shock region.


We believe that the strong X-ray flux produced in the ring collision
is the dominant source of energy input to the ejecta.  The X-ray flux
from the ring increased by a factor $\sim 3$ in the 0.5 -- 10~keV band
between day $6,000$ and day $8,000$\cite{Racusin2009}, similar to what
we find for the optical emission from the ejecta. To investigate this,
we calculated the fraction of X-rays absorbed by the ejecta from a
point source located at the ring, using the partially mixed 14E1
explosion model\cite{Blinnikov2000}. As shown in Supplementary Fig.~6,
most of the observed X-ray flux is absorbed in the core region of the
ejecta (corresponding to speeds less than $5,000\ \kms$), where
most of the heavy elements reside.  At an energy of $\sim 0.35$ keV,
which corresponds to the temperature of the dominant component in the
X-ray spectrum\cite{Zhekov2010}, the fraction of flux absorbed by the
ejecta at $t_{\rm yr}$ years can be approximated by $1.6 \times
10^{-3} t^{1.67}_{\rm yr}$ (the increase with time is mainly due to
the increasing solid angle of the expanding ejecta, assumed to be
spherical, as seen from the ring). This gives a present-day absorbed
X-ray luminosity of $ \sim 5.0 \times 10^{35} \ergs$. In this
calculation we have neglected the weaker, hard component that
contributes to the X-ray spectrum\cite{Zhekov2010}. We note that this
does not significantly affect the estimate of the absorbed flux,
although the hard X-rays may be important due to their larger
penetrating power.

To model the ejecta light curve produced by input from the X-rays we
scaled the observed X-ray flux\cite{Racusin2009} by the fraction
absorbed at 0.35 keV, multiplied the resulting flux with a constant
(corresponding to the conversion efficiency from X-rays to optical)
and added this to the radioactive energy input. Fig.~\ref{fig4} shows
the scaled X-ray flux together with the observed light curves. This
model follows the general trend of the observed fluxes in both bands,
although we note that a more accurate model would need to take into
account the detailed shape of the X-ray spectrum and the reprocessing
of the X-rays into optical emission. The required conversion
efficiency from X-rays to optical emission in our model is $5.0\%$ in
the R-band and $3.1 \%$ in the B-band.

The conversion of X-rays to optical/IR emission is similar to that of
the ${}^{44}$Ti positrons. Both involve degradation of non-thermal
electrons to heating, ionization and excitation. For a typical
ionization fraction of $10^{-3} - 10^{-2}$, the expected efficiency of
conversion from X-rays to H$\alpha$ (the dominant line in the R-band)
is $\sim 5 \%$ (Supplementary Information, section 3). This conversion
factor is consistent with the scaling factor we used to model the
light curve. Similar arguments apply to the B-band.  Furthermore, the
density in the core is high enough for the time-scale of recombination
to be shorter than the expansion time-scale, ensuring a balance
between the energy input and output.

Other possible explanations for the increase in flux include input
from a central pulsar\cite{Woosley1989}, a transition from optically
thick to optically thin dust and positron leakage from the Fe-rich
regions. We find that input from a pulsar is unlikely for several
reasons. In particular, it would be a strange coincidence for the
emission from the pulsar to mimic the increasing X-ray flux from the
ring interaction. Also, we expect the energy input from a pulsar to be
concentrated toward the low-velocity material at the centre of the
ejecta, but observations of the H$\alpha$ and [Ca II] lines show
that the increase occurs for speeds up to $\sim 5,000$ \kms. We
also note that constraints on a point source at the centre of SN 1987A
have already been obtained using HST data taken near the minimum of
the ejecta light curve\cite{Graves2005}. A change in the properties of
the dust or a transition in the positron deposition process are also
unable to quantitatively explain the observed increase in flux
(Supplementary Information, section 4).


We conclude that SN~1987A has made a transition from a radioactively
dominated phase to a phase dominated by the X-ray input from the ring
collision. This conclusion has interesting implications for the
observed morphology. In particular, most of the X-rays are likely to
be absorbed at the boundary of the ejecta core, where the density
increases rapidly. This may lead to the light from the ejecta being
emitted in a ring that is concentrated to the plane of the
circumstellar ring. The 'hole' in the ejecta (Fig.~\ref{fig1}), which
has become more pronounced since $\sim 2001$, may in fact be a result
of this, rather than reflecting the true density distribution or dust
obscuration. The asymmetric morphology seen at speeds of $\la 3,000$
\kms~ in the near-IR [Si I] and [Fe II] lines\cite{Kjaer2010} is,
however, likely to be intrinsic to the metal core.  By studying future
changes in the morphology of the ejecta, we will be able to understand
the origin of this asymmetry

In the future the density of the ejecta will decrease and the fraction
of X-rays absorbed will grow (Supplementary Fig.~7). As a result
the ionization will increase and a smaller fraction of the X-ray flux
will produce line excitation. A larger fraction will go into
heating, leading to an increase in the mid-IR flux and a flattening of
the optical light curves. In time, the X-rays will also penetrate
deeper layers of the ejecta, thereby allowing us to probe the chemical
structure of the innermost ejecta. This will be a novel form of X-ray
tomography.

\subsection*{Supplementary Information}

Supplementary information accompanies this paper.

\subsection*{Acknowledgments}

This work was supported by the Swedish Research Council and the
Swedish National Space Board.  Support for the HST observing program
was provided by NASA through a grant from the Space Telescope Science
Institute, which is operated by the Association of Universities for
Research in Astronomy, Inc.

\subsection*{Author contribution}
J.L. carried out the data reduction and analysis together with G.\"O.,
P.G., B. L., J.S. and P.C.; C.F. performed the theoretical modelling
together with A.J. and C.K.; J.L. and C.F. wrote the paper. R.P.K. is
PI for the HST/SAINTS collaboration. All authors discussed the results
and commented on the manuscript.

\subsection*{Author information}
Reprints and permissions information is available at
www.nature.com/reprints. The authors declare no competing financial
interests. Readers are welcome to comment on the online version of
this article at www.nature.com/nature. Correspondence and requests for
materials should be directed to J.L. (josefin.larsson@astro.su.se) or
C.F. (claes@astro.su.se).

\newpage

\subsection*{Figures}

\begin{figure}[h!]
\begin{center}
\includegraphics{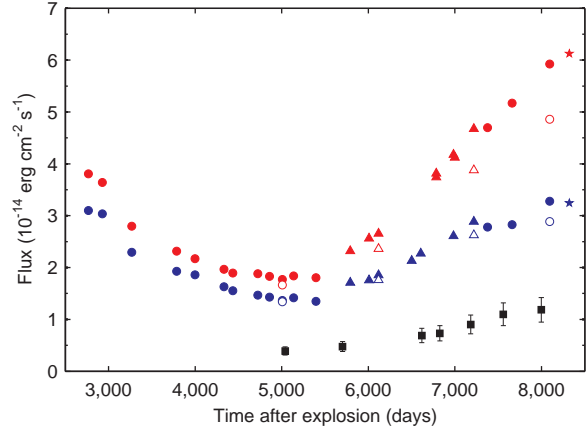}
\caption{Light curves of the ejecta in different wavebands. Data
  points from WFPC2, ACS and WFC3 are shown as dots, triangles and
  stars, respectively. Red and blue symbols correspond to the R- and
  B-bands. A correction factor has been applied to the ACS and WFC3
  fluxes in order to account for the differences between these
  instruments and WFPC2. To quantify the contamination from the
  brightening circumstellar ring we created detailed models of the
  ring for a number of different epochs (Supplementary Information,
  section 1, and Supplementary Figs.~3 and 4). The open symbols show
  the ejecta fluxes after the contribution from the ring has been
  removed. Although the contamination from the ring increases with
  time it never exceeds $\sim 20 \%$ of the flux from the ejecta. The
  statistical errors on the HST fluxes are smaller than the data
  points and systematic errors are discussed in the Supplementary
  Information, section 1. The black squares show the flux of the [Ca
    II] \wll 7291.5, 7323.9 lines from VLT/UVES (error bars,
  s.d.). These lines are free from contamination from the
  circumstellar ring and the reverse shock. We note the decreasing flux
  during the first $\sim 5,000$ days and the increase thereafter,
  indicating an extra source of energy.  A further indication that the
  energy source has changed is that the colour, determined from the
  flux ratio between the B-and R-bands, changes from a level of $\sim
  0.8$ up to day $5,000$ to a value close to $\sim 0.6$ on day
  $8,000$.
\label{fig2}}
\end{center}
\end{figure}
\begin{figure}[h!]
\begin{center}
\includegraphics{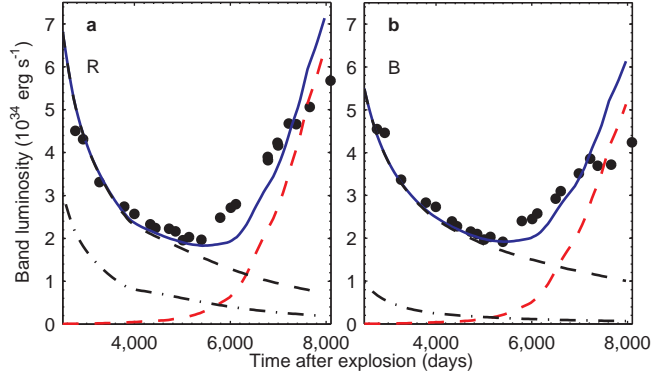}
\caption{Evolution of the luminosity from the ejecta in the R- and
  B-bands. The dashed black lines show a model with only radioactive
  input, mainly from ${}^{44}$Ti. The ${}^{44}$Ti mass used for the
  model is $1.4 \times 10^{-4}$ \msun \cite{Jerkstrand2011}, as
  determined from a detailed model that takes into account the effects
  of scattering and fluorescence of the UV flux, as well as a
  correction for internal dust (here taken to have a 50\% covering
  factor of the core inside 2000 \kms)\cite{Lucy1991,Wooden1993}. The
  lower dot-dashed lines show the light curves with no ${}^{44}$Ti,
  illustrating the necessity to include this isotope. The red dashed
  lines show a model based on a constant fraction of the observed
  X-ray flux\cite{Racusin2009}, corrected for the fraction of flux
  absorbed by the ejecta (Supplementary Fig.~7). The blue solid line
  shows the sum of both models.  The R- and B-bands contain 5\% and
  3\% of the total bolometric flux, respectively, and we expect these
  fractions to remain roughly constant with time. This is because the
  relative amount of energy resulting in heating, ionization and
  excitation will remain nearly constant as long as the ionization by
  the X-rays is $\la 10^{-2}$.  It is clear from this figure that
  there is a transition at $\sim 5,000$ days from a radioactivity
  dominated phase to a phase dominated by the X-ray input from the
  collision with the ring.
\label{fig4}}
\end{center}
\end{figure}

\end{document}




\noindent
 {\Large {\bf {\fontfamily{phv}\selectfont Supplementary information }}}

\section{Photometry}

Our analysis includes all HST imaging observations performed between
1994 and 2009 in the the red (WFPC2/F675W, ACS/F625W and WFC3/F675W)
and blue (WFPC2/F439W, ACS/F435W and WFC3/F438W) broad-band filters
(the two groups of filters will be referred to as the R- and B-bands
from here on). The images were drizzled to combine dithered exposures
and to remove cosmic rays. Details of the observations are given in
Supplementary Table~1 and a selection of images are shown in
Supplementary~Fig.~\ref{sample}. The images show the expansion of the
asymmetric ejecta and the brightening hot spots in the circumstellar
ring.

We measured the brightness of the ejecta as a function of time using
the IRAF/STSDAS task ELLIPSE, using different elliptical and circular
apertures described below. The measurements give mean flux densities,
which we convert into total fluxes by multiplying with the rectangular
bandwidth of the different filters. When converting to total fluxes we
also correct for the fact that there are some differences between the
filters used for the three instruments. To determine the corrections
we used the HST FOS spectrum from 1997$^{22}$ (this covers the entire
spectral range probed by all the filters) and used the SYNPHOT package
to calculate flux densities in the different filters.  After
converting to total fluxes we obtain the following flux ratios:
(ACS/F625W) / (WFPC2/F675W)$\ = 1.08$, (WFC3/F625W) /
(WFPC2/F675W)$\ = 1.16$, (ACS/F435W) / (WFPC2/F439W)$\ = 1.27$ and
(WFC3/F438W) / (WFPC2/F439W)$\ = 1.06$. In all the light curves that
we show we have corrected the ACS and WFC3 fluxes in accordance with
this. A small offset still remains (see e.g.  Fig. 2), which may be
due to charge transfer efficiency losses in WFPC2 (see discussion
below). Changes in the spectral energy distribution (SED) would also
have some effect on the conversion factors, although we note that
observations$^{22,}$\cite{Wang1996} suggest that the SED does not vary
significantly with time.

For each of the observations we also determined the sky brightness
using the IRAF/FITSKY routine. Specifically, we determined the
centroid sky value in five empty regions located around SN~1987A and
then used the average of these measurements to subtract the sky
background from the light curve. The background is typically only a
few per cent of the intrinsic ejecta brightness. We also considered
the possibility of contamination from the two bright stars that are
located to the north-east and south-west of SN~1987A. These stars have
strong diffraction spikes from the secondary-mirror spiders, which may
pass through the ejecta depending on the HST roll angle. We inspected
all the images and found that there is only one observation
(2006-04-15) where there is a strong spike passing through our
apertures. We estimated the contribution from this spike by measuring
the brightness of the other spikes from the same star, and corrected
the light curves accordingly. Another source of contamination is one
of the narrow outer rings of SN 1987A, which is partly projected on
top of the ejecta. However, the small, decreasing
flux\cite{Tziamtzis2010} of this component is negligible compared to
the emission from the ejecta.

The shape of the ejecta is roughly elliptical so as a first step we
used an elliptical aperture to measure the evolution of the flux. The
shape of the ellipse was determined by fitting elliptical isophotes to
the ejecta region in the observation from 2000-11-13. This epoch was
chosen because it is long enough after the explosion that the ejecta
shape is clearly discernible, while still early enough for the edge of
the inner ejecta to be clearly separated from the circumstellar
ring. The fits were performed using the IRAF/STSDAS task
ELLIPSE. Setting the semi-major axis (sma) to $0.2''$ (the size used
for the light curve in Fig.~2 for this epoch) we find a position angle
of $17.4 \pm 0.1 \dg$ and an eccentricity of $0.684 \pm 0.002$. This
can be compared with previous measurements with HST (P.A.~$= 14 \pm
5\dg$)\cite{Wang2002} and VLT/SINFONI (P.A.~$= 15 \pm 0.9\dg$,
$e=0.83$, from the $1.644~\mu \mathrm{m}$ Si~I/Fe~II feature)$^{21}$.
The agreement with the position angle is clearly good and, given that
the SINFONI measurements probe the Si/Fe core and the R-band is
dominated by the H-rich gas, the agreement with the eccentricity is
also reasonable. As described in the main paper, the light curves
presented in Fig.~2 were derived using this elliptical aperture. The
size of the aperture was expanded as a function of time in order to
always follow the same ejecta material.

In order to make sure that the transition from a decaying to a rising
flux seen in Fig.~2 is not connected to the use of this particular
aperture, we also performed the measurements using apertures with
other shapes and sizes, trying both apertures that have a fixed size
and apertures that expand with the
ejecta. Supplementary~Fig.~\ref{apertures} shows the R- and B-band
light curves for the same expanding elliptical aperture as in Fig.~2,
an expanding circular aperture (with radius equal to the sma of the
elliptical aperture) as well as a large, fixed aperture. The fixed
aperture encloses most of the ejecta but still avoids direct
contamination by the ring (the aperture is shown in Fig.~1). The light
curves extracted from the expanding elliptical and circular apertures
show very similar behaviour, demonstrating that the exact shape of the
aperture is not important. The light curve from the large, fixed
aperture also shows the same behaviour during the first $\sim$ 7,000
days, but levels off at late times. This is at least partly due to
ejecta material expanding out of the fixed aperture. The large
aperture is also likely to be significantly affected by contamination
from the circumstellar ring, making it hard to draw firm conclusions
regarding the ejecta using this aperture.

Contamination from the ring (due to flux from the brightening ring
spilling into the ejecta region) is likely to affect all our light
curves at some level. In order to quantify this effect we used the HST
point-spread-function modelling tool Tiny Tim\cite{Tinytim} and
created models for the ring for a number of different epochs in both
bands. We set up the models to match the observed distribution of
hot spots as closely as possible and to have the same total flux in
the ring area as the observations. The modelling is illustrated in
Supplementary~Fig.~\ref{tinytim}, which shows the real data and the
model for the F675W image from 2009-04-29, and in
Supplementary~Fig.~\ref{tinytim_int}, which compares the radial
profiles of the data and the model. The corrected light curves derived
using these models are shown as open symbols in Fig.~2 and
Supplementary~Fig.~\ref{apertures}. The contribution from the ring
increases with time and is larger in the R-band, but it does not have
a major effect on the light curves from the expanding apertures.  For
the large, fixed aperture the contamination is much more severe,
reaching about 50\% of the total flux at late times in the R-band.

The modelling of the ring is naturally associated with some
uncertainties. In order to quantify these uncertainties we performed
two different tests. First, we investigated the effect of varying the
number of points sources used to model the ring. Specifically, we
created models using between 8 and 90 point sources distributed along
the ring, while always keeping the total flux in the ring area
constant. The different ring models resulted in fluxes in the ejecta
apertures that agreed to within 15\%. This demonstrates that the
specific model for the ring is not very important as long as the same
total flux is distributed in an ellipse around the ejecta. The main
reason for this is of course that we are interested in the total flux
that leaks into the ejecta region and not its detailed distribution.

Another test of the models is to compare the flux in the synthetic and
real images in regions located just outside the circumstellar ring. We
used clean regions (uncontaminated by stars and other sources of
emission) that allow us to measure the flux in the wings of the point
sources in the ring. For the observation from 2009-04-29 we find that
the agreement is excellent (the discrepancy is less than 10\%), which
gives us confidence that the models provide a good estimate of the
emission from the ring. For earlier epochs it is harder to evaluate
the models in this way since the clean regions outside the ring are
completely dominated by the sky background rather than ring
emission. For this reason we cannot determine whether our models are
equally accurate at early times. We note, however, that the flux from
the ring is negligible at these times.

An additional source of uncertainty is the loss of Charge Transfer
Efficiency (CTE). This effect is known to increase with time and is
especially important for faint objects on low backgrounds located far
from the detector readout\cite{Dolphin09}. To investigate the
importance of CTE losses we performed aperture photometry on about 10
stars of varying brightness located around SN~1987A. We found that the
effect is negligible in the ACS but clearly noticeable in WFPC2, where
stars lose at most $\sim 10 \%$ of their flux between 1994 and
2009. This also agrees with estimates obtained for the supernova
ejecta using the WFPC2 CTE correction formula\cite{Dolphin09}. In
terms of the light curve in Fig.~2, we note that a $\sim 10
\%$ increase in the WFPC2 data points at late times would improve the
agreement between the ACS and WFPC2.

\section{Spectroscopy}
\label{spectra}

As a complement to the photometry we investigate HST STIS spectra
taken with the G750L grating in 2004 and 2010. These spectra overlap
with the wavelength region covered by the R-filters and give us an
opportunity to study the different spectral components that may
contribute to the increase in flux. In both observations a slit with a
width of 0.2$''$ was placed in the north-south direction, centred on
the ejecta$^{16,}$\cite{Heng2007}. In 2004 there were also two
observations taken with the narrow slit placed on either side of the
central slit. To allow for a comparison with the photometry we
extracted spectra in regions that match the size of the expanding
elliptical aperture as closely as possible. This means using the
central slit only and an extraction height of $\pm 0.24''$ in 2004 and
$\pm 0.31''$ in 2010.

Supplementary~Fig.~\ref{stis} shows the H$\alpha$ profiles from 2004
and 2010. The line profiles are clearly asymmetric with a strong blue
wing extending down to $\sim -5,000\ \mathrm{km\ s^{-1}}$ and a red
wing that only reaches velocities of about $\sim
2,500\ \mathrm{km\ s^{-1}}$.  The constant width of the ejecta line
profile shows that the expanding aperture that we used follows the
same material over a 6-year span of time. From the two-dimensional
spectra$^{16,}$\cite{Heng2007} it is clear that the blue wing
mainly comes from the northern part of the ejecta. The lack of a
correspondingly strong red wing may be due to dust absorption in the
far side of the ejecta. Supplementary~Fig. \ref{stis} also shows that
the increase in flux between 2004 and 2010 is mainly due to an
increase in the blue wing of the ejecta profile.

The H$\alpha$ emission also contains a small contribution from the
reverse shock. This component is clearly visible on the red side of
the line as the boxy part of the profile between $\sim 3,000-
9,000\ \mathrm{km\ s^{-1}}$. On the blue side of the line the reverse
shock appears at velocities higher than $\sim -
9,000\ \mathrm{km\ s^{-1}}$, where it is blended with the broad [O I]
$\lambda\lambda$ 6300, 6364 lines from the ejecta at $-
12,022\ \mathrm{km\ s^{-1}}$ and $- 9,096\ \mathrm{km\ s^{-1}}$. To
estimate the flux from the reverse shock we focus on the boxy part of
the profile on the red side, since this is not blended with any other
strong lines. The contribution from this component compared to the
total flux in the R-band is $\sim 9-10 \%$ in both 2004 and 2010. It
is clear from inspection of the spectra that the reverse shock is not
as strong on the blue side, and a very conservative upper limit on the
total contribution from the reverse shock is therefore $\sim 20
\%$. We especially note that there is no significant increase between
2004 and 2010.

In Supplementary~Fig.~\ref{stis} we also show the spectrum obtained by
combining the three slit positions from the 2004 observations,
renormalized to the continuum of the other spectra. The combined
spectrum is very similar to that from the central slit, albeit with a
slightly stronger line core and red wing. The contribution from the
reverse shock on the red side is $\sim 11 \%$, i.e.~only marginally
larger than the values given above.

The STIS observation from 2010 also allows us to investigate the
emission from the southernmost part of the ejecta, which is seen
outside both the expanding and the fixed aperture at late times
(Fig.~1). STIS spectra extracted in this region show a significant
contribution from the reverse shock but also an asymmetric line
profile from the ejecta, with a strong blue wing extending down to
$\sim -2,000\ \mathrm{km\ s^{-1}}$. The fact that we see blueshifted
emission from the southern part of the ejecta shows that there must be
some ejecta material located well above the plane of the ring.

The [Ca II] observations in Fig.~2 were performed between 2000 and
2009 with the Very Large Telescope (VLT) at ESO, using the UVES
instrument. The primary purpose of these observations was to follow
the evolution of the narrow lines from the ring collision. To maximise
the spectral resolution and to minimise contamination from nearby
stars a comparatively narrow slit, 0.8 \arcsec~wide, was chosen. The
seeing was in most cases 0.6 -- 0.8 \arcsec, which is just enough to
spatially resolve the ring into a northern and a southern part. The
slit was centred on the position of the supernova and had a position
angle of 30$^{\circ}$ in all the observations. The observations were
reduced with the standard pipeline and the errors on the absolute
fluxes were estimated to be less than 20 -- 30\% (ref.~5).

\section{Modelling of the light curve}

At first, the energy radiated by a SN comes from the shock wave
arriving at the surface of the exploding star. This phase lasted
lasted only a few hours in SN~1987A, and was followed by a period of a
few weeks where the light curve was powered by the diffusive release
of the internal energy produced by the shock
wave$^{17}$. Subsequently, the energy for emission was supplied by the
decay of radioactive isotopes created in the explosion. During the
first $\sim1,000$ days the radioactive input was dominated by
${}^{56}$Ni, decaying into ${}^{56}$Co and ${}^{56}$Fe. From the first
two years of the bolometric light curve the mass of ${}^{56}$Ni could
be determined to $0.069$~\msun\cite{Bouchet1991}. After $\sim 1,000$
days ${}^{57}$Ni took over and a mass of $3 \times 10^{-3}$ \msun
\ was found\cite{Kurfess1992,Varani1990,Fransson1993}. Finally, after
$\sim 1,600$ days the next most abundant isotope, ${}^{44}$Ti, has
been predicted to dominate the energy input to the ejecta. While the
bolometric light curve could be accurately determined up to $\sim
1,000$ days, the dominance of the mid-IR, which is difficult to
determine from observations, requires accurate modelling of the
optical bands at later times. In addition, the hydrogen-rich parts of
the ejecta are subject to a freeze-out\cite{Fransson1993}, where
recombination and cooling become slow compared to the radioactive
decay time-scales and the expansion time-scales. It is therefore
necessary to use a time-dependent approach for the modelling.

The modelling of the radioactively powered part of the light curves in
Fig.~3 is based on a time-dependent spectral
code\cite{Kozma1998I,Kozma1998II}.  The code includes calculations of
the non-thermal excitation and ionization by the gamma-rays and
positrons from the radioactive decay of ${}^{56}$Ni, ${}^{57}$Ni and
${}^{44}$Ti, as well as a full non-local thermodynamic equilibrium
treatment of the most important elements and lines. The present
version also includes updated atomic data, in particular more accurate
collision rates for Fe II, and other ions. The radiative transfer uses
the Sobolev approximation, which treats the line escape in individual
lines accurately. However, the non-local scattering from line to line
is not included (except for the Ca II H and K lines, which feed the
[Ca II] \wll 7291.5, 7323.9 lines by fluorescence). This especially
affects the UV and blue parts of the spectrum and leads to an
underestimate of the flux in the B-band$^{12}$.

An important test is that this model accurately reproduces the
H$\alpha$ evolution up to day 3,600 (ref.~2). A comparison with the
spectrum at 7.8 years$^{22}$ shows agreement to within 50\% for the
R-band, but the model underestimates the flux in the B-band by a
factor of 2.5. To model the shape of the light curves we therefore
renormalize the B-band fluxes by a factor 2.5 and the R-band by a
factor 1.5. The total energy input from radioactivity can be estimated
to $\sim 1.2 \e{36} \ergs$ at 20 years$^{12}$. In addition to this
there is a substantial contribution from delayed recombination
(freeze-out). This comes primarily from the low-density hydrogen
envelope and is seen mainly as Balmer emission. The decrease of this
component is substantially faster than the ${}^{44}$Ti decay and is
responsible for the fast decline seen before $\sim 4,000$ days
(Fig.~3).

When translating the broad-band fluxes to luminosities we assume a
distance of 50 kpc and a reddening of E$_{\rm B-V} =
0.16$~mag\cite{1990AJFitzpatrick}. For the comparison with the
observations we use the light curve from the expanding elliptical
aperture, multiplied by a constant factor for the R- and B-bands. The
reason for using this, rather than the fixed aperture in Fig.~1, is
that the latter shows a misleading flattening at epochs later than
$\sim 7,000$ days because of the expansion of part of the ejecta
outside the aperture. The constant factors were determined as an
average of the ratio of the flux in the fixed aperture to that of the
expanding aperture for epochs earlier than $7,000$ days. The constants
are 2.6 for the R-band and 2.5 for the B-band. There is of course some
uncertainty in the correction from the expanding aperture to the total
emitted flux, which also affects the ${}^{44}$Ti mass.

To determine the X-ray input to the ejecta we calculated the
photoelectric absorption along different rays in the ejecta from a
point source located at the radius of the circumstellar ring. For the
abundance and density distributions we used the partially mixed 14E1
explosion model$^{17}$. This model is spherically symmetric (although
mixed). Unfortunately, no 2-D or 3-D simulations exist with
appropriate nucleosynthesis. The spherical approximation should,
however, be sufficient for the qualitative nature of the model. Most
of the envelope is transparent to X-rays with energies above about
0.2~keV as seen in Supplementary~Fig.~\ref{tau1}, which shows the
ejecta expansion velocity for optical depth $\tau = 1$ as a function
of photon energy. The high metallicity core dominates the opacity by
K- and L-shell absorption by oxygen and other elements, and is opaque
at all but the very highest energies.

The energy deposition is an angular average of the absorbed intensity
over all directions of the core and envelope, shown in
Supplementary~Fig.~\ref{fig3c} for four different photon energies. The
different deposition curves can be well approximated by $C(E) \times
10^{-3} t^{1.67}_{\rm yr}$, where $C(E) = (4.5, 1.6, 0.7)\times
10^{-3}$ for $E =$ 0.1, 0.35, and 1 keV, respectively. The general
time dependence can be explained as a balance between the increase of
the solid angle of the core as seen from the ring, $\propto t^2$, and
the increasing transparency of the ejecta as the density decreases.

The X-ray luminosity used in this paper is that directly observed by
Chandra$^{6}$. This is likely to be an underestimate for several
reasons. The quoted X-ray luminosities only refer to the observed flux
$\ga 0.5$ keV. For a shock with a temperature of 0.35~keV most of the
luminosity is below this energy, especially if the shock is radiative,
as is likely for shocks with velocities below $\sim
400\ \kms $ (ref.~3). In addition, the line profiles of the optical
emission show that there are shocks of lower velocities, most likely
tangential shocks\cite{2002Pun}, which radiate mainly below the
observable X-ray band and add to the ionizing flux. The typical shock
velocity from the optical lines is $\sim 200\ \kms$ (ref.~5), which
corresponds to a temperature of $\sim 50$ eV. Because the opacity
increases as $E^{-3}$ (Where $E$ is the photon energy), the absorption
efficiency of the ejecta of this low energy radiation is nearly
100\%. The absorbed X-rays are therefore more than enough to explain
the observed increase in the optical bands. A more accurate estimate
of the spectral distribution would require a detailed numerical
calculation of the X-ray absorption and thermalization, which is
beyond the scope of this paper. We also remark that both the optical
emission from the ring$^{5}$ and the mid-IR$^{8}$ are strongly
correlated with the X-rays.

The conversion from X-rays to optical emission takes place first by
photoelectric absorption, creating fast non-thermal electrons with
energies up to keV. These are thermalized by excitation, ionization
and heating of the free, thermal electrons. For hydrogen at the
relevant degree of ionization, $35 - 40\%$ of the secondary electron
energy goes into ionization, followed by recombination, while $\sim
8\%$ goes into direct excitation of $n > 2$ (refs 39, 40). At
$100-200$ K, $\sim 9$\% of the ionization energy and $\sim 16\%$ of
the excitation energy emerges as H$\alpha$ (ref. 41). In total this
corresponds to $\sim 5\%$ of the X-ray energy. Other elements add to
this. The relative importance of excitation, ionization and heating
will remain roughly constant as long as the ionization is $\la
10^{-2}$, and we therefore do not expect conversion efficiency for the
two bands to change over time. However, once the ionization surpasses
this value, the fraction going into heating increases at the expense
of the other channels. This will increase the far-IR fine structure
lines compared to the optical lines until the temperature is high
enough for thermal effects to increase the optical/UV
emission\cite{Kozma1998I,Kozma1998II}.

\section{Effects of dust and positron leakage in the ejecta}

 There is strong evidence for dust in the ejecta from early
 observations$^{23,24,}$\cite{Moseley1989}, as well as from
 observations discussed in this supplement. In principle, a transition
 from optically thick to optically thin dust could account for an
 increase in the flux. This is, however, unlikely for several
 reasons. First, it would be a strange coincidence that this would
 occur at the same time as the X-ray flux becomes
 important. Secondly, compared to the flux from our model with
 constant covering fraction (i.e. optically thick dust clumps), which
 is given by the dashed line in Fig.~3, the increase has to be a
 factor of $\sim 8$ in the R-band and $\sim 5$ in the B-band. The
 estimated covering factor of the dust at early times was $\sim
 0.5$ (refs 23, 24), implying a maximum increase in the flux by a factor
 of $\sim 2$. Moreover, we expect the increase to occur first in the
 R-band and later in the B-band, based on the higher extinction in the
 latter band. The imaging also indicates dust obscuration in the
 central part of the SN up to the latest epochs
 (Supplementary~Fig. \ref{sample}), although this may also be an
 artefact of the external X-ray illumination, as discussed in the main
 paper.

Finally, positron leakage from the Fe-rich material containing the
${}^{44}$Ti may enhance the emission from the O- and H-rich zones in
the core$^{12}$. This depends sensitively on the strength of a
magnetic field in the ejecta, which is uncertain. The main effect of
positron leakage is that the hydrogen and intermediate element lines
would increase at the expense of the Fe I-II lines. At most this may
enhance the emission in the observed bands by a factor of $\sim 2$
compared to the complete trapping case. As discussed above, this is
far from sufficient to explain the observed increase in the R- and
B-bands.


\newpage
\section{Supplementary Figures}

\begin{figure}[h!]
\begin{center}
\resizebox{160mm}{!}{\includegraphics{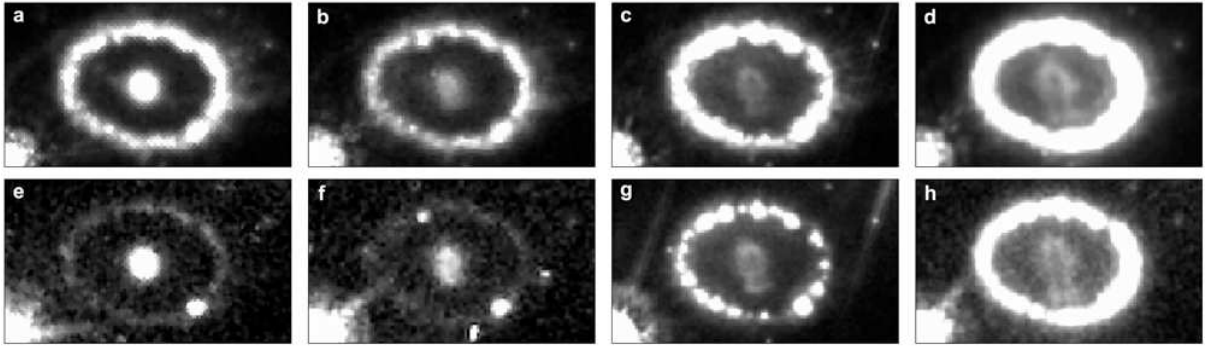}}
\caption{Selection of HST images of SN~1987A. The images show the
  evolution of the ejecta and the circumstellar ring as a function of
  time. The top row shows the R-band and the bottom row shows the
  B-band. The observing dates are 1994-09-24 ({\bf a} and {\bf e}),
  1999-04-21 ({\bf b} and {\bf f}) , 2003-11-28 ({\bf c} and {\bf g})
  and 2009-04-09 ({\bf d} and {\bf h}), which correspond to 2,770,
  4,440, 6,122 and 8,101 days after the explosion, respectively. Note the
  expansion of the ejecta and the brightening hot spots in the
  circumstellar ring.
\label{sample}}
\end{center}
\end{figure}

\begin{figure}
\begin{center}
 \resizebox{82mm}{!}{\includegraphics{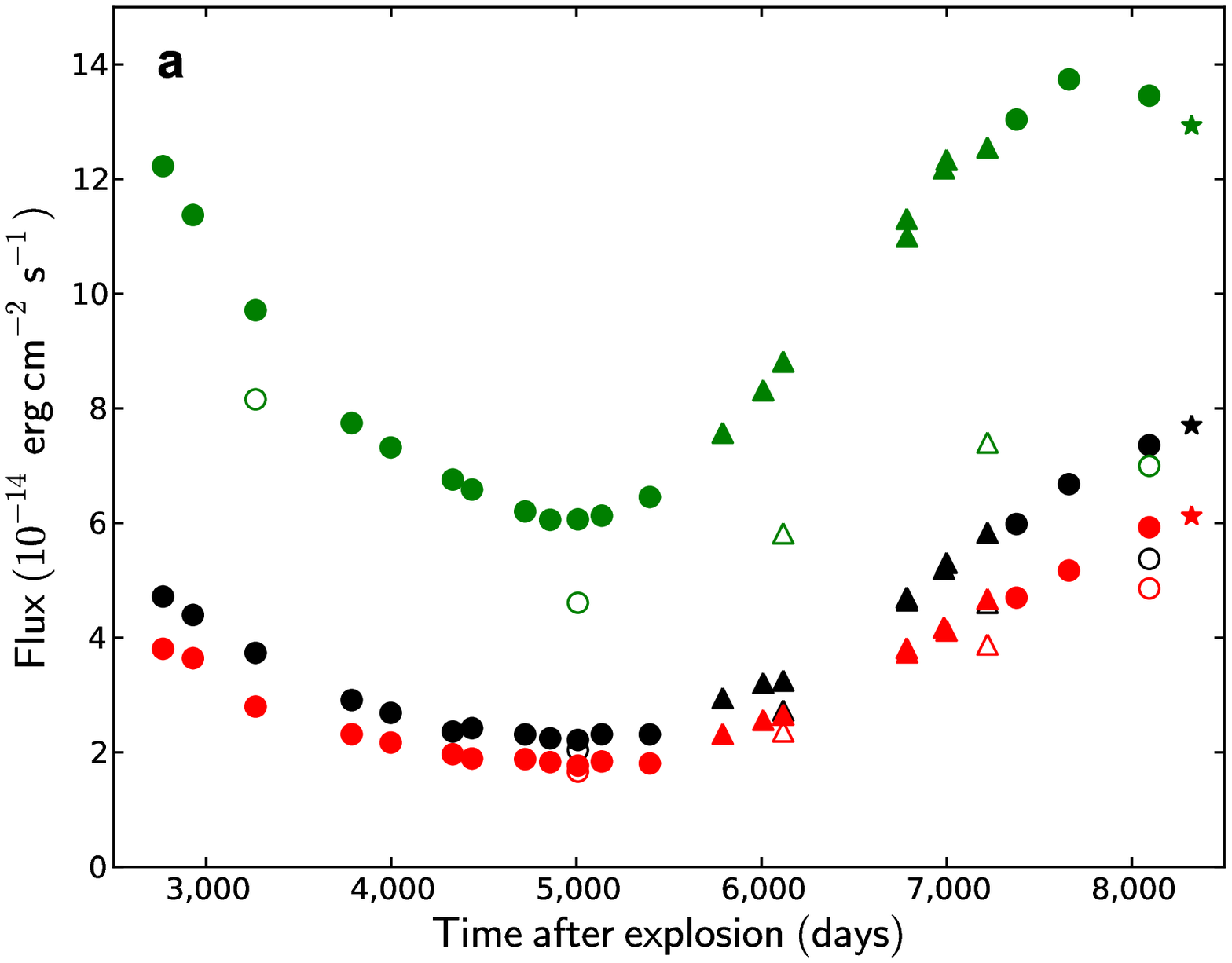}}
  \resizebox{82mm}{!}{\includegraphics{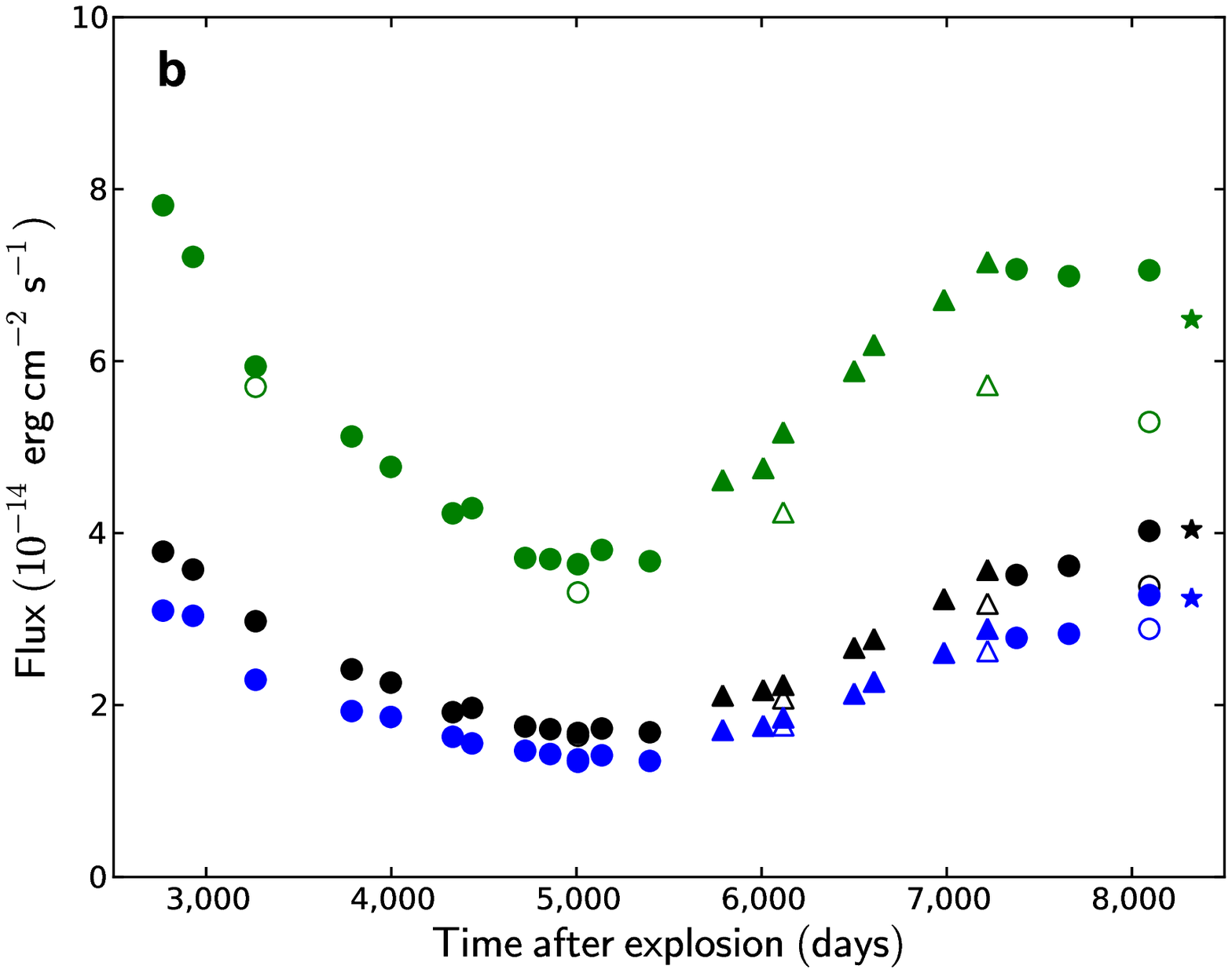}}
\caption{Comparison of ejecta light curves for different
  apertures. Panel {\bf a} shows the R-band and panel {\bf b} shows
  the B-band (note the different scales on the y-axes). Data points
  from WFPC2, ACS and WFC3 are shown as dots, triangles and stars,
  respectively. A correction factor has been applied to the ACS and
  WFC3 fluxes in order to account for the differences between these
  instruments and WFPC2. The red and blue data points correspond to
  the same expanding elliptical aperture as in Fig.~2, shown here for
  reference. Green data points are for a fixed aperture that encloses
  as much of the ejecta as possible (sma $=0.50''$, see also Fig.~1)
  while black data points are for an expanding circular aperture with
  radius equal to the semi-major axis of the expanding elliptical
  aperture. Open symbols indicate approximate flux levels after
  accounting for emission from the circumstellar ring. The light curve
  from the fixed aperture clearly levels off at late times. This is at
  least partly due to ejecta material expanding out of the aperture,
  but other effects may also contribute.
\label{apertures}}
\end{center}
\end{figure}

\begin{figure}
\begin{center}
\resizebox{82mm}{!}{\includegraphics{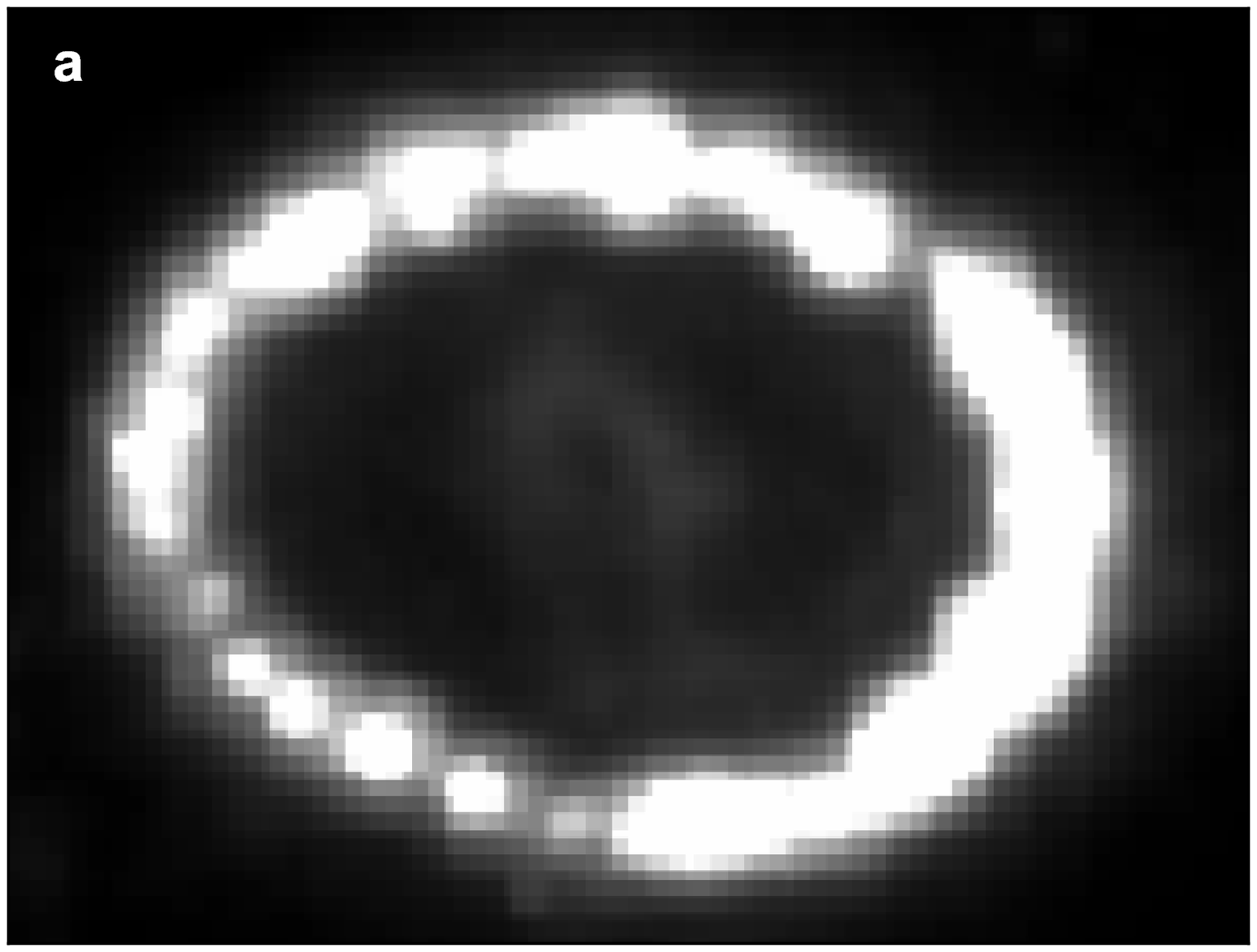}}
  \resizebox{82mm}{!}{\includegraphics{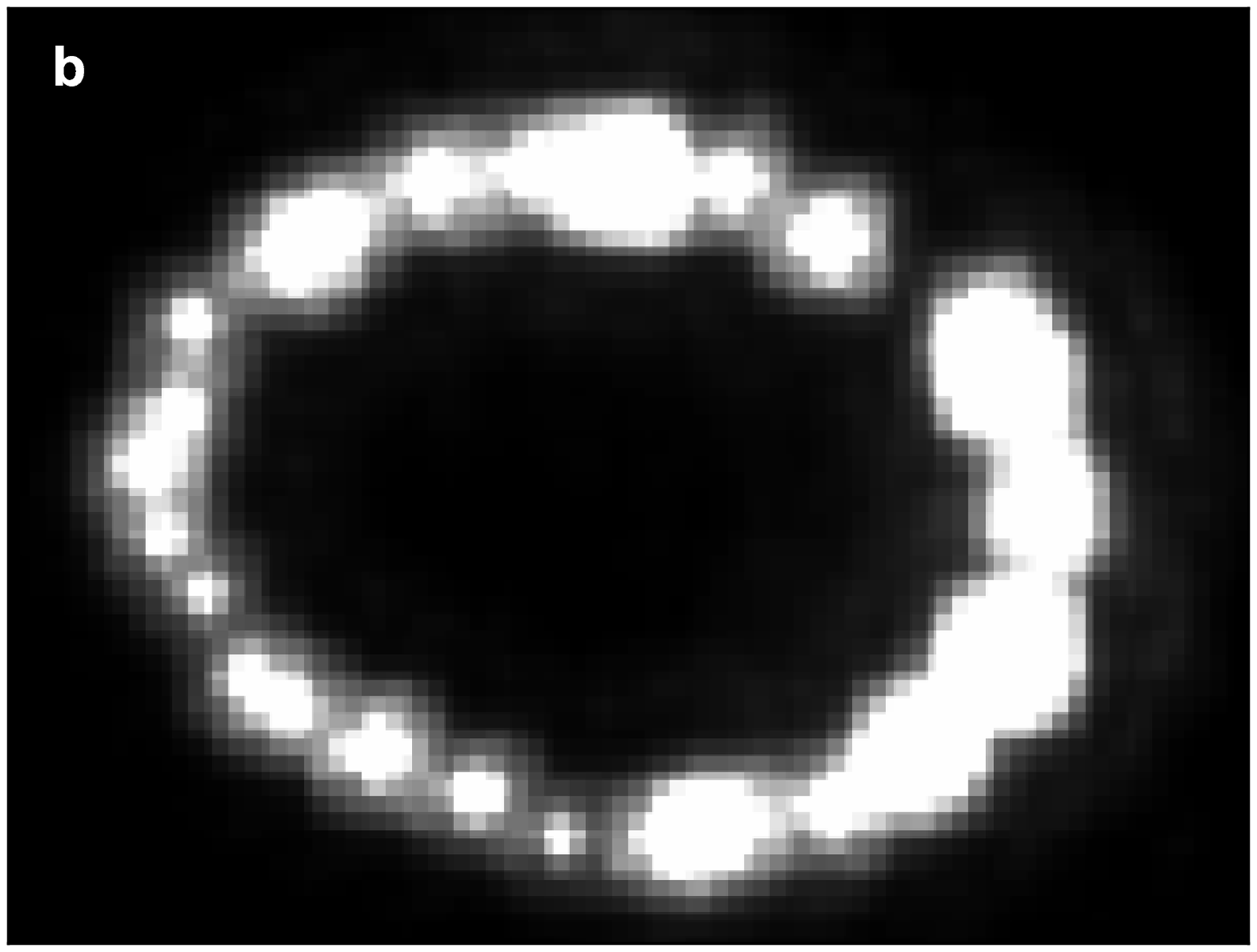}}
\caption{Example of the modelling of the ring. {\bf a}, The HST/WFPC2
  F675W image from 2009-04-29. This is the same image as in
  Supplementary Fig.~1d but the image levels have been adjusted to
  show the structure of the ring more clearly. As a result the ejecta
  appear very faint. {\bf b}, The model for the hot spots in the ring for
  the same epoch.
\label{tinytim}}
\end{center}
\end{figure}

\begin{figure}
\begin{center}
   \resizebox{120mm}{!}{\includegraphics{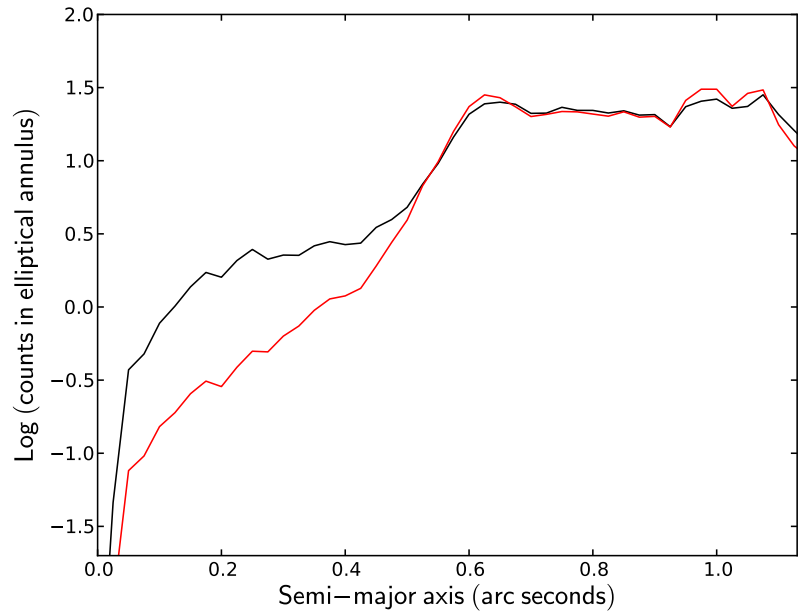}}
\caption{Comparison of the radial brightness profiles of the real data
  and the model. The profiles were produced for the 2009-04-29
  WFPC2/F675W observation (Supplementary~Fig.~\ref{tinytim}), with
  the real data shown in black and the model in red. Specifically the
  lines show the counts inside elliptical annuli aligned with the
  circumstellar ring as a function of the sma of the annulus. The
  model clearly agrees well with the real data in the ring area
  (between about 0.6--1.0$''$) and should therefore give us a good
  estimate of the amount of flux from the ring that leaks into the
  ejecta region.
\label{tinytim_int}}
\end{center}
\end{figure}

\begin{figure}
\begin{center}
\resizebox{120mm}{!}{\includegraphics{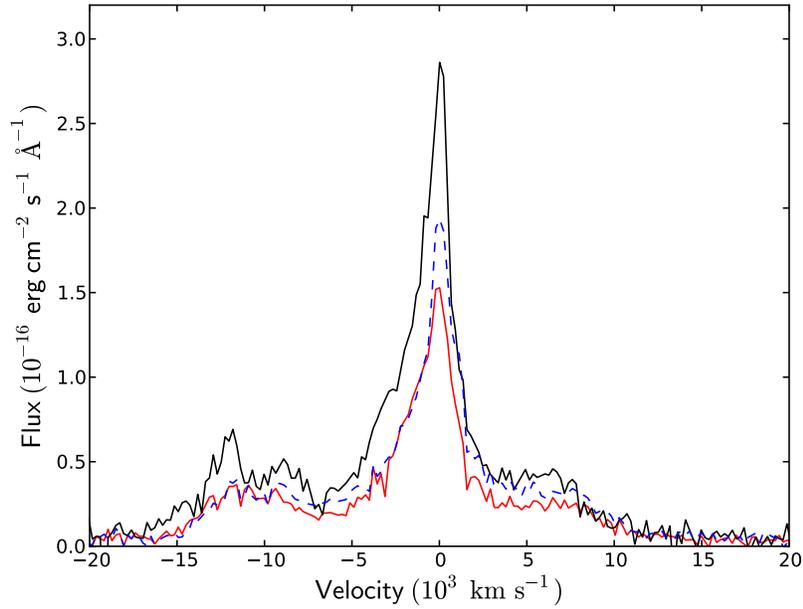}}
\caption{STIS H$\alpha$ profiles of the ejecta. The red and black
  lines are for a 0.2$''$ slit placed in the north-south direction and
  centred on the ejecta in 2004 and 2010, respectively. The dashed
  blue line is the combined spectrum from three 0.2$''$ slits placed
  next to each other in 2004, renormalized to match the continuum
  level of the other spectra. The spectral extraction regions were
  chosen to correspond as closely as possible to the size of the
  expanding elliptical aperture used for the flux measurements in
  Fig. 2. The boxy part of the line profile between $3,000$ and
  $9,000\ \mathrm{km\ s^{-1}}$ is due to the reverse shock. There is
  also a contribution from the reverse shock on the blue side, but
  this is harder to disentangle, because of blending with the [O I]
  $\lambda \lambda\ 6300, 6364$ lines at $-
  12,022\ \mathrm{km\ s^{-1}}$ and $- 9,096\ \mathrm{km\ s^{-1}}$.
\label{stis}}
\end{center}
\end{figure}

\begin{figure}[h!]
\begin{center}
\resizebox{120mm}{!}{\includegraphics{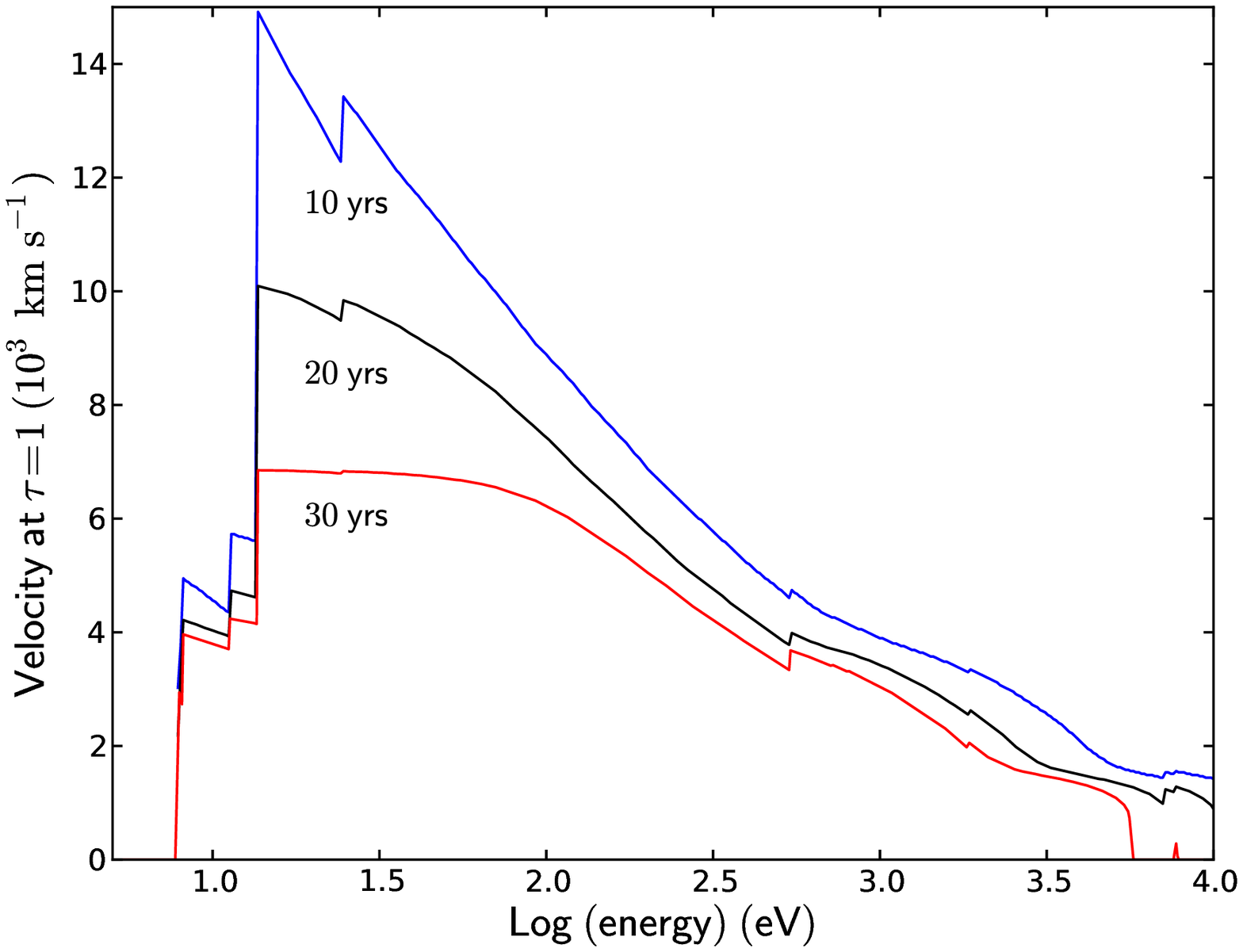}}
\caption{Ejecta velocity corresponding to optical depth $\tau=1$.  The
  optical depth is measured from the ring as a function of energy at
  10, 20 and 30 years after the explosion. The maximum velocity for
  each of these curves is given by the ejecta velocity in the plane of
  the ring, which is $2.1 \e4 (t/10 \ {\rm yrs})^{-1}$ \kms. Note the
  relatively small change in velocity with time of $\tau(E) = 1$ for
  energies $\ga 0.2$ keV. The FWHM of the line profiles are therefore
  expected to decrease relatively slowly with time, as long as most of
  the shock energy is emitted above this energy. Note also that at 30
  years the full ejecta will be transparent at $\ga 6$ keV.
\label{tau1}}
\end{center}
\end{figure}

\begin{figure}
\begin{center}
\resizebox{120mm}{!}{\includegraphics{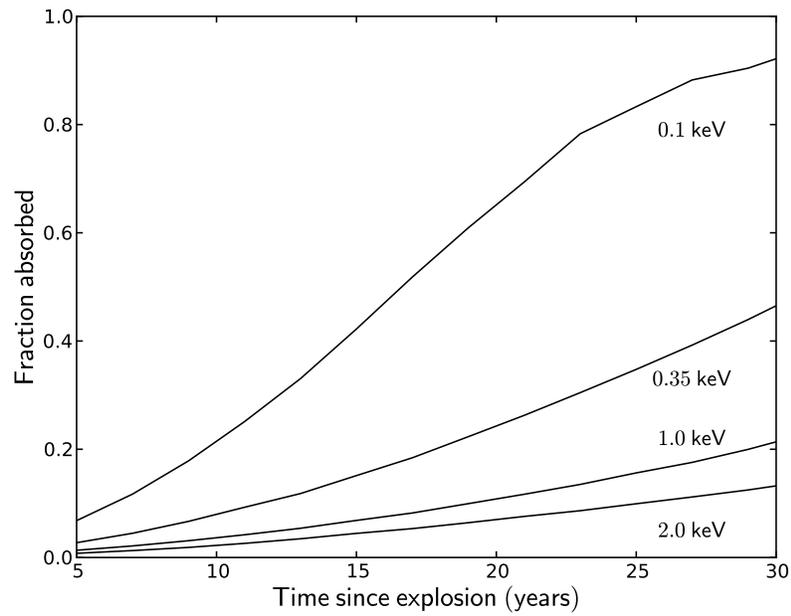}}
\caption{Fraction of the X-ray flux absorbed by the supernova ejecta
  as a function of time. The different lines correspond to different
  photon energies as indicated in the figure. The main reason for the
  evolution is the increasing solid angle of the core as seen from the
  ring ($\propto t^2$), damped by the increasing transparency at
  higher energy.
\label{fig3c}}
\end{center}
\end{figure}

\clearpage

\section{Supplementary Table}

\spacing{1}

\begin{table}[h!]
\begin{center}
\begin{tabular}{lrcrc}
\hline\hline \\ [-8pt]
\multicolumn{1}{c}{Date} & \multicolumn{1}{c}{Instrument / Filter}
& \multicolumn{1}{c}{Exposure time} & \multicolumn{1}{c}{Instrument / Filter} 
&  \multicolumn{1}{c} {Exposure time} \\
 &    & \multicolumn{1}{c}{(s)} &    & \multicolumn{1}{c}{(s)}   \\ \hline
1994-09-24 & WFPC2 / F675W  & 600 & WFPC2 / F439W & 800  \\

1995-03-05 &  WFPC2 / F675W & 600 & WFPC2 / F439W & 800  \\

1996-02-06 &  WFPC2 / F675W & 600 & WFPC2 / F439W & 950  \\

1997-07-10  &  WFPC2 / F675W  & 600 & WFPC2 / F439W & 800  \\

1998-02-06  &  WFPC2 / F675W  & 400 & WFPC2 / F439W & 500  \\

1999-01-07  &  WFPC2 / F675W  & 1220 & WFPC2 / F439W & 1039  \\

1999-04-21  &  WFPC2 / F675W  & 400 & WFPC2 / F439W & 800  \\

2000-02-02  &  WFPC2 / F675W  & 400 &  WFPC2 / F439W & 460  \\

2000-06-16  &  WFPC2 / F675W  & 400 & WFPC2 / F439W & 800  \\

2000-11-13  &  WFPC2 / F675W  & 2400 & WFPC2 / F439W & 1200  \\

2001-03-23  &   WFPC2 / F675W & 500 &  WFPC2 / F439W & 1000 \\

2001-12-07  &  WFPC2 / F675W  & 800 &  WFPC2 / F439W & 1600  \\

2003-01-05  &  ACS / F625W  & 800 &  ACS / F435W & 1200 \\ 

2003-08-12  &  ACS / F625W  & 480 &  ACS / F435W & 800 \\ 

2003-11-28  &  ACS / F625W  & 800 &  ACS / F435W & 1600 \\

2004-12-15  &    &  &  ACS / F435W & 1600 \\

2005-04-02  &     &    &   ACS / F435W & 1200 \\

2005-09-26  &  ACS / F625W  & 12000  &    & \\

2005-09-28  &  ACS / F625W  & 720 &    & \\

2006-04-15  &  ACS / F625W  & 1200 &   ACS / F435W & 1200 \\

2006-04-29  &  ACS / F625W  & 720 &    & \\

2006-12-06  &  ACS / F625W  & 1200 &    ACS / F435W & 1800 \\

2007-05-12  &   WFPC2 / F675W  & 2700 &  WFPC2 / F439W & 3000 \\

2008-02-19  &   WFPC2 / F675W & 1600 &  WFPC2 / F439W & 2400 \\

2009-04-29  &   WFPC2 / F675W  & 1600 &  WFPC2 / F439W & 2000 \\

2009-12-12  &  WFC3 / F625W  & 3000 &  WFC3 / F438W & 800 \\

\hline
\end{tabular}
\caption{\label{obstab}\small{HST observations used for the
    analysis.}}
\end{center}
\end{table}

\spacing{2}

\clearpage
\newpage

\bibliography{hst_ejecta}
\bibliographystyle{naturemag}



\clearpage